\documentclass[aps,prb,twocolumn,showpacs,superscriptaddress]{revtex4}
\usepackage{epsfig}
\usepackage{times}
\usepackage{amsmath}
\begin{document}

\title{Oscillons: an encounter with dynamical chaos in 1953?}

\author{S. Denisov}
\email[Electronic address (corresponding author): ]{sergey.denisov@physik.uni-augsburg.de}
\author{A. V. Ponomarev}
\affiliation{Institute of Physics, University of Augsburg,
Universit\"{a}tstr.~1, D-86159 Augsburg}
\date{\today}
\begin{abstract}
We discuss the works of one of electronic art pioneers,  Ben F. Laposky (1914-2000),  and argue that he might have been the first  to create a family of essentially nonlinear analog circuits that allowed him to observe chaotic attractors.
\end{abstract}


\maketitle

\textbf{As any branch of science, dynamical chaos theory has no definite birthdate or starting point, but a certain number of milestones. Edward N. Lorenz recognized the chaotic attractor, which now bears his name, in  numerical simulations somewhere around 1961 \cite{Lorenz}, while Leon Chua incarnated chaotic attractors with a specially designed electronic circuit in 1983 \cite{Chua}. Here we present evidences that Ben F. Laposky  was able to create a family of analog circuits that allowed him to observe chaotic attractors and other trademarks of nonlinear science as early as 1953.}

Although Laposky, a draftsman by profession, had received a proper recognition as a pioneer of electronic art,
at no time his name has emerged in the context of dynamical chaos theory. The circuits he had implemented for generation of ``oscillons'' on the screen of a cathode ray tube oscilloscope, remain a mystery \cite{justify}. It is known  that some of his thirty-seven circuits \cite{lap1} had ``as many as 70 different setting of controls''\cite{lap2} and that ac-voltage has been used for the circuit feeding.  Our analysis is based on the vanity press booklet with the still photos of the fifty-six oscillons, which were exhibited at the Sanford Museum (Cherokee, Iowa) in 1953 \cite{lap1}.

There are three  oscillons that captured our attention (Figs.~1, A-C). The first oscillon (Fig.~1A,  $\#  30$ in the booklet) looks very similar to the celebrated R\"{o}ssler attractor \cite{ross} (Fig.~1D), while the second one (Fig.~2B, $\#  6$) looks like a multi (in this case, three)- scroll attractor \cite{scroll} (Fig.~1E),  a typical output of chaotic circuits which include nonlinear elements with  nonmonotonous input/output characteristics.

\begin{figure}[t]
\center
\includegraphics[width=0.48\textwidth]{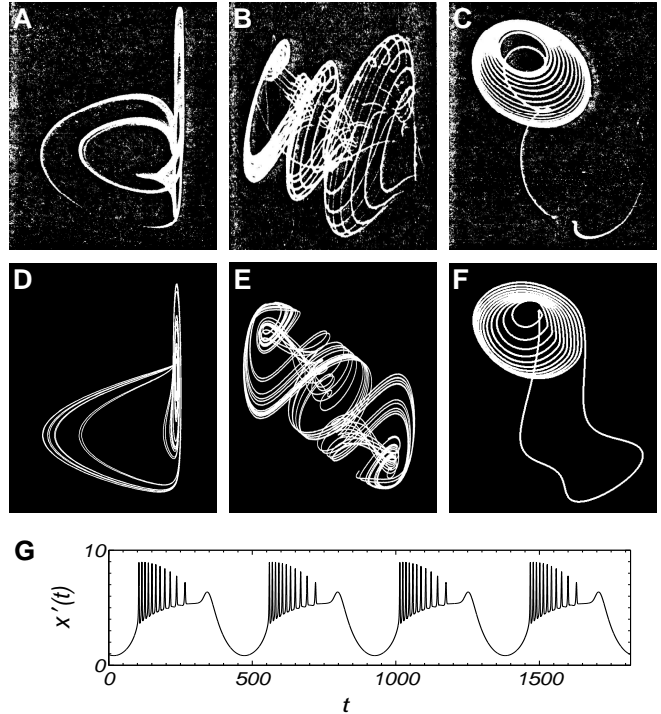}
\caption {Oscillons from the booklet \cite{lap1}, (A) $\#  30$; (B) $\#  6$; and (C) $\#  36$.
 R\"{o}ssler attractor, with the parameters $a=0.2$, $b=0.19$, and $c=7.1$ (D), three-scroll attractor, $\dot{x}=y$, $\dot{y}=z$, $\dot{z}=-z-5y-x^5+4x$ \cite{sprott1} (E), and limit-cycle regime of the Hindmarsch-Rose model (F).  While the last attractor was modified by a nonlinear frame transformation, the first two were obtained by a simple adjusting rotation of the original frame $(x,y,z)$ and subsequent projection on the $(x',y')$-plane. (G) The periodic bursting pattern, $x'(t)$, generated by the limit cycle from Fig.~1C, bottom row.} \label{fig4}
\end{figure}

The words ``looks similar'' stand here for the similarity of the attractor shapes only. Obviously, this observation cannot be taken as a rigorous proof of the statement that Laposky observed chaotic attractors, yet it certainly confirms the fact that the electronic systems he used were essentially nonlinear. The third oscillon (Fig.~1C, $\#  36$) provides more substantial evidence of the last statement. It represents  a peculiar limit cycle, which, we believe, demonstrates the bursting phenomenon \cite{burst}, a dynamical regime of neuronal activity where a neuron periodically fires discrete series of spikes, and which can be reproduced with nonlinear models only. In order to substantiate this claim we employed the Hindmarsh-Rose  model \cite{neuron1}, with the parameters $x_R=-1.6, r=0.01, s=4$, $a=2.5$, and $I=3$. Using the shape of the model attractor in a three-dimensional space $(x,y,z)$, we tried to find a frame transformation, $\mathbf{F}=\{ F_x(x,y,z), F_y(x,y,z), F_z(x,y,z)\}$, which maps the original frame space onto a new one, $\mathbf{F}:(x,y,z) \rightarrow (x', y', z')$, so that the new coordinates of reference attractor points, $(x_{r}', y_{r}')$, would maximally approach the corresponding reference points of the oscillon, $(X_{r}, Y_{r})$ (see Appendix A).   Namely, we first  digitized the image of the oscillon and distributed twenty-six reference points over it. Since allowable frame transformations should be smooth in order to preserve  the topology of the attractor, we assumed the functions $F_x, F_y, F_z$ to be forth-order polynomials, and then minimized the quantity $\Delta= \sum_{r} (x_{r}' - X_{r})^{2} + (y_{r}' - Y_{r})^{2}$  in a standard least-squares fashion,  by varying  the polynomial coefficients. The resulting limit cycle  (Fig.~1F) has the shape of  the original oscillon, and produces a typical periodic bursting pattern (Fig.~1G). Thus the oscillon $\#  36$ might be a fingerprint of an analog model of neuron, which was created even before the very first mathematical model was presented \cite{neuron2}. Finally, we would like to underline that the trajectory shown on Fig. 1C exhibits a specific type of instability \cite{burst} which is irreproducible with simple linear setups (for example, by crumpling a Lissajous curve).

Our results give substantial evidences that Ben F. Laposky had all  ingredients needed to encounter chaotic regimes of analog electronic systems and it is quite \textit{probable} that he had witnessed these regimes while tuning the parameters of his mysterious circuits \cite{altgough}. Yet ``science starts from problems, and not from observations'' \cite{poper}, so that, even if our hypothesis is correct, Laposky cannot be nominated for the discovery of chaotic attractors (although he certainly deserved to be mentioned in the curriculum vitae of electronic chaos; see Appendix B).
But what is most exciting about the oscillon story  is that the use of pure aesthetic criteria, which guide artist's preferences, had led to the selection of several aperiodic, chaotic attractor-like structures from more than 6,000 images \cite{lap1} -- decades before scientists started to talk about the ``beauty of chaos'' \cite{beat}.

\section{Appendix A}
\textbf{Notes:} We have used the standard routine from the Matlab Optimization Toolbox for the solution of the optimization problem \cite{opt}.

\begin{center}
\includegraphics[width=0.48\textwidth]{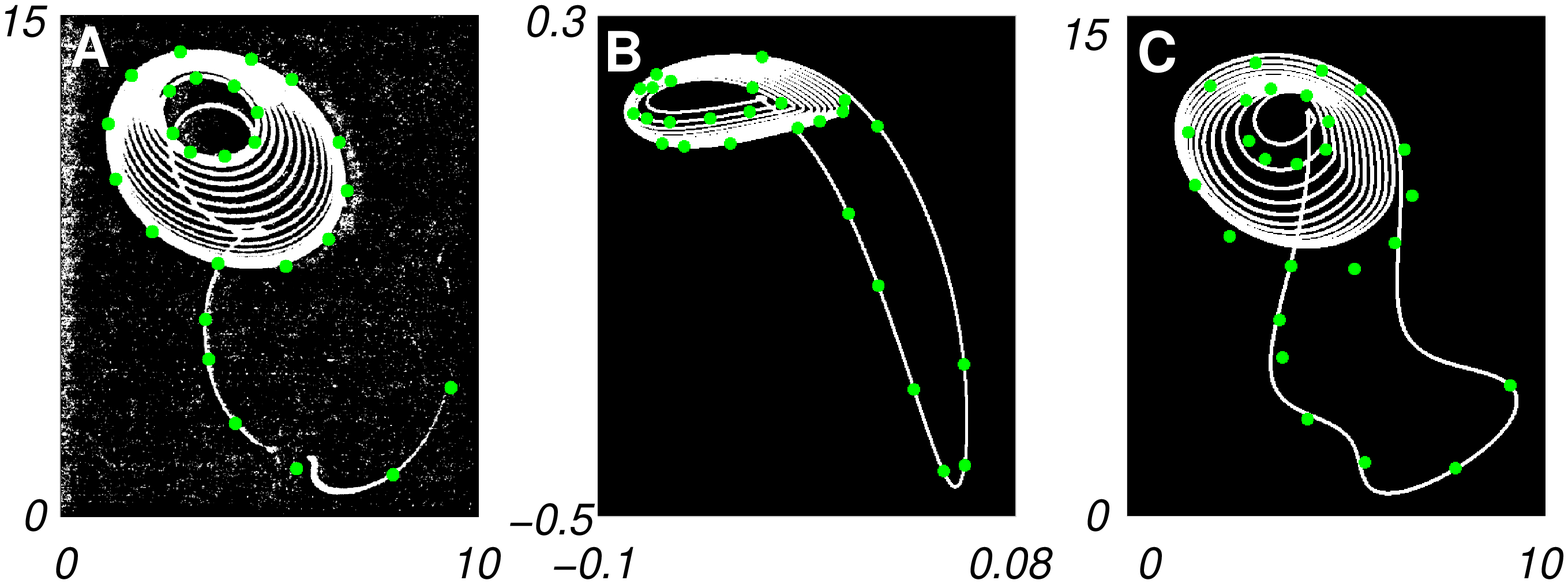}
\label{S1}
\end{center}
{\small FIG.~S1: \textbf{Phase-space transformation.} (A) Oscillon $\#  36$ from Ref.\cite{lap1} with twenty-six reference points $(X_r, Y_r)$, (B) original limit cycle of the Hindmarsch-Rose model with reference points $(x_r, y_r)$; (C) limit cycle after a frame transformation with superimposed reference points $(X_r, Y_r)$.}
\\

\section{Appendix B}
\textbf{Electronic chaos prehistory:}
\begin{itemize}
 \item $^\star$\underline{1927} First encounter with electronic chaos: ``Often an irregular noise is heard... [that] strongly reminds one of the tunes of a bagpipe''
 [B. van der Pole and J. van der Mark, Nature \textbf{120}, 363 (1927)];

 \item \underline{1953} {\it Ben  F. Laposky's oscillons} \cite{lap1,lap2};\\
 
 \item \underline{1957} First encounter with dynamical chaos in numerical studies(?): Two-disc dynamo model by Tsuneji Rikitake: ``...the system in question performs an extremely complicated oscillation.''
       [T. Rikitake, Proc. Cambridge Philos. Soc. \textbf{54}, 89 (1958)] (see also \cite{rik});\\

 \item $^\star$\underline{1958} Experiments with glow-lamp ring circuits:
       ``Larger rings [of glow lamps], on the other hand, perform
       erratically and are always hard to adjust for a specific firing order.''
       [R. L. Ives, Electronics \textbf{31}, 108 (1958)];\\
 \item \textbf{\underline{around 1961} E. Lorenz found his attractor: the rise of Chaology}  \cite{Lorenz};
  \item $^\star$\underline{1961, November 16th} Chaotic signal from an overloaded traveling wave tube amplifier:
``High power noise source employing a feedback path around a traveling wave tube''
	[C. A. Reis \& J. E, Zellens, Pat. USA Number: 3,176,655 filed Nov. 16, 1961];
 \item \underline{1961, November 27th} While experimenting with vacuum tube circuits,  Yoshishuke Ueda noticed
       ``randomly transitional phenomena'' (though his finding was not published until 1970 \cite{ueda}).

 \item $^\star$\underline{1965} An overloaded amplifier with a feedback element generates aperiodic broad band oscillations  [E. A. Koptyrev \& L. E. Pliss, Radiotekhnika i elektronika \textbf{10}, 1828 (1965) (in Russian)];
        \item ....................
        
 \item \underline{1979} Experiments with a self-excited circuit:
       `` ... the results of a theoretical, numerical and experimental investigation of one of the simplest self-excited noise generators. ... the statistical properties of the signal are determined... by the intrinsic dynamics of the system rather than by.. the noise''
       [S. V. Kiyashko, A. S. Pikovskii and M. I. Rabinovich, Radio Eng. and Electron. Phys. \textbf{25}, 74 (1980)];\\
        
 \item \textbf{\underline{1983} Chua's circuit}: incarnation of chaotic attractor regimes with a specially designed circuit \cite{Chua}.
 \end{itemize}

$^\star$ All these researchers did not attribute the observed phenomena to the underlying chaotic dynamics of the employed systems. They interpreted chaotic signals as the result of amplification of ``inevitable" internal fluctuations.


\begin{thebibliography}{}
\bibitem{justify} We didn't find any information concerning circuits used by Laposky (circuit diagrams, elements, regimes). We doubt that such information exsists and is aviable at all.
\bibitem{altgough} Strickly speaking, we should not exclude the possibility that, by feeding the vertical and horizontal control inputs of an oscilloscope with specially devised quasiperiodic signals, one could get something similar to the oscillons shown on Figs. A, B.
\end{thebibliography}
\end{document}